# INTEGRATING LABVIEW INTO A DISTRIBUTED COMPUTING ENVIRONMENT

K.U. Kasemir, M.Pieck, L.R. Dalesio, LANL, Los Alamos, NM 87544, USA

Abstract

Being easy to learn and well suited for a self-contained desktop laboratory setup, many casual programmers prefer to use the National Instruments LabVIEW environment to develop their logic. An ActiveX interface is presented that allows integration into a plant-wide distributed environment based on the Experimental Physics and Industrial Control System (EPICS). This paper discusses the design decisions and provides performance information, especially considering requirements for the Spallation Neutron Source (SNS) diagnostics system.

## 1 INTRODUCTION

EPICS is a highly configurable toolset for building distributed control systems that scale to accommodate large projects[1]. It has C and C++ interfaces for the integration of new hardware and software[2], full source code is available.

While this provides the best performance, highest flexibility and is easily understood by experienced programmers, the initial EPICS setup does already require a network connection and two computers: the real time target and a Unix or Win32 (Windows NT, 9x, 2000) host. Some application engineers who are unfamiliar with the multifaceted EPICS toolset prefer to start with a purely visual environment on a single PC. LabVIEW is a tool where the engineer most familiar with the application task can quickly start the implementation. This paper presents ways of integrating LabVIEW into the distributed EPICS environment.

## 2 EPICS CHANNEL ACCESS

EPICS communicates via the ChannelAccess (CA) network protocol. Front-end input/output controllers (IOCs) run a CA server, presenting values as well as time stamps, limits, units, alarm status and other attributes. CA clients locate the server based on channel names. They establish a connection and subscribe to changes in value or connection status. Management of the connection status as well as high throughput are key features of CA [3,4].

CA server and client libraries are available to C/C++ software on Unix and Win32. Since the CA libraries need to monitor network connections for incoming requests and data, the user program has to implement periodic calls into the CA libraries. The CA server monitors a dedicated UDP port for search requests. It is therefore suggested to run only one CA server per computer since an additional server would use a non-standard UDP port, unknown to most CA clients.

## 3 LABVIEW EXTENSION OPTIONS

LabVIEW can call Win32 DLLs, communicate via ActiveX and DDE or perform low-level UDP/TCP network calls. Using the latter would result in a re-write of the CA libraries. Since LabVIEW code is unlikely to compete with a C/C++ implementation, this was not attempted. Using the CA libraries as DLLs is problematic because LabVIEW would have to initiate the periodic network processing.

DDE has been used for the CA client library: A separate program implemented the periodic processing of CA network connections, presenting the data to MathWorks MATLAB via DDE[2]. LabVIEW can use this DDE interface, but DDE is deprecated with the advent of newer technologies, namely Win32 COM (Component Object Model) and ActiveX[5].

## 4 ACTIVEX CA SUPPORT

Two ActiveX Automation Server programs interface to the CA server respectively client library. Every COM-aware Win32 program can create an 'EpicsCAServer.ProcessVariable' or 'EpicsCAClient.ProcessVariable'. COM marshals requests from different processes into a single thread, avoiding threading problems in the CA library. More than one program can transparently use the same instance of the CA server or client. LabVIEW, MATLAB, Microsoft Visual Basic and Visual C++ offer interactive browsing of COM objects, modification of published properties ('name', 'units', etc.), invocation of methods ('setValue', …) and reaction to events ('Changed', 'NewValue').

Several LabVIEW VIs shield the user from the underlying COM calls. Serving a number is reduced to one initial call to a "Create" VI that takes the name of the new process variable, followed by calls to a "Set" VI whenever the value changes, see Fig. 1. In this example LabVIEW serves a read-only process variable to EPICS clients. Fig. 2 shows a more realistic setup as suitable for a *setpoint*, a variable that is to be changed both locally on the LabVIEW front panel and remotely via a CA client. After creating a process variable for the setpoint, additional informational parameters are configured and then the value of a user interface knob on the front panel is served. In addition, LabVIEW polls for input from CA and modifies the value of the knob in response.

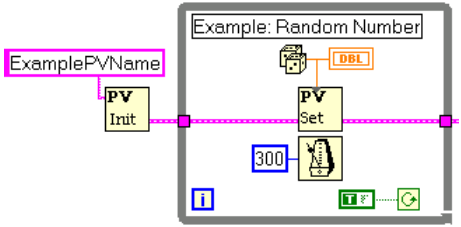

Figure 1: Serving a random number from LabVIEW

While other languages can asynchronously react to ActiveX events, invoking callbacks immediately after the event arrives, LabVIEW offers only a polling or waiting mechanism to check for events.

## 5 PERFORMANCE

The COM call to update the value of one process variable requires 0.14 milliseconds for LabVIEW on a 900 MHz Pentium PC, increasing with the data size. Times for Visual Basic, a compiled language, are slightly better (Table 1). Repeated measurements showed variations of up to 15% on a Windows NT 4.0 PC because neither LabVIEW nor Win32 are deterministic. The CPU load was at 100% in these tests, leaving no time for the CA server to actually respond to client requests. In a realistic setup delays will be needed to allow for CA client interaction.

The measured times reflect the overhead of COM calls. They also apply when the CA server sends an event to LabVIEW. Every time an operator changes a setpoint on an EPICS operator screen, LabVIEW has to receive this value (1 COM call), maybe constrict it to the allowed operating range and post the result to the server (1 COM call), resulting in an expected overhead of 0.28 ms per value.

Table 1: Times for updating data on the ActiveX CA server

| Data Served | LabVIEW | Visual Basic |
|---|---|---|
| Double | 0.14 ms | 0.08 ms |
| Double[100] | 0.20 ms | 0.16 ms |
| Double[500] | 0.45 ms | 0.40 ms |
| Double[1000] | 0.75 ms | 0.77 ms |

Scaled linearly, one could serve around 700 values at 10 Hz. In reality, different timings result depending on the implementation. As an example, handling 10 setpoint variables in a loop required 7 ms, 100 variables required 70 ms. An alternative parallel implementation handled 10 setpoints in only 0.50 ms. While faster than the loop, this is impractical for many setpoints because the resulting LabVIEW diagram is indecipherable.

The Low Energy Demonstration Accelerator (LEDA) at the Los Alamos Neutron Science Center has several operational LabVIEW systems. One handles 10 power supplies for 52 magnets, 8 ion pumps, 3 ion gauges, 3 beam line valves plus 40 thermocouples. The readbacks and status values result in a channel count of about 525, there are ~175 setpoints for outputs and interlock limits. LabVIEW polls for user input between handling the hardware, reaction times are 1-2 seconds. In another LEDA system LabVIEW controls 20 power supplies via GPIB, resulting in ~140 channels. LabVIEW has to generate the GPIB commands and then wait for a response. Though the sequence of sending and receiving GPIB messages has been optimized, the reaction to user input is 3s or more, still acceptable for the specific system.

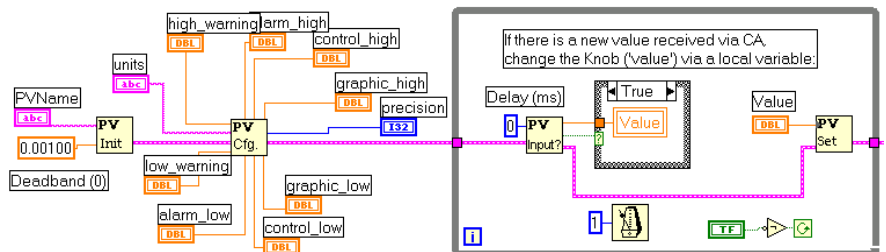

Figure 2: Serving a setpoint from the LabVIEW front panel, responding to CA input.

## 6 SNS DIAGNOSTICS ISSUES

SNS diagnostics systems like the beam position monitors (BPM) require handling of up to 100 setpoints while sending the measured beam parameters at 10 Hz. We assume four values. Since the beam is pulsed at 60 Hz, this higher update rate is desirable at certain times. All data is to be time-stamped according to information sent on the real-time data link (RTDL). Occasionally, measurements are to be taken in response to an event link signal. The system shall respond to user input within 1s. On demand, array information about a full beam pulse of 2500 samples shall be provided, but since this is allowed to cause delays we will ignore it in the following discussion.

With LabVIEW updating four values at 60 Hz and handling 100 setpoints, at least 0.9s of each second are left for processing the diagnostics hardware. Limited by polled operation, LabVIEW cannot asynchronously respond to event-link signals or retrieve the current time stamp in a deterministic manner. A solution is to monitor the event link and RTDL in hardware, time-stamp the data in hardware and have LabVIEW only read the result. The ActiveX CA Server was recently extended so that LabVIEW can pass these time-stamped values.

Since the planned diagnostics hardware is for the PCI bus and LabVIEW cannot directly access it, a Win32 device driver is required, written in C or C++. Its implementation might be simplified by basing it on commercially available real-time extension software for Win32.

Past experience has shown problems related to setpoints. In a conventional EPICS IOC, they are simply named and response to user input is instantaneous. A LabVIEW program has to check for user input and react to it. When this is done in sequence with the remaining program tasks, delays of several seconds have been observed. While LabVIEW does offer multithreading for handling this in parallel, the arising threading issues require advanced LabVIEW training.

Another proposal for the SNS diagnostics is to keep the diagnostics systems minimal, resulting in one PC per BPM running LabVIEW, and use an ordinary EPICS IOC to collect and correlate data across BPMs to provide consistent information for beam orbits.

## CONCLUSION

We presented a way of integrating LabVIEW and other Win32 programs (MATLAB, Visual Basic) into a distributed EPICS environment. The ActiveX interface to CA is easy to learn. It is ideal for small LabVIEW systems, especially temporary setups like beam-line experiments. It is used successfully at LEDA for operational systems with several hundred process variables.

The current performance measurements together with recent enhancements suggested that LabVIEW, integrated via ActiveX, could meet SNS requirements, although this is neither a pure nor a simple LabVIEW implementation.

A new system like the SNS BPMs naturally involves both new hardware and software. For an EPICS IOC, the required BPM driver, a C program, would handle the data from several BPMs, using the same event and RTDL support as other SNS IOCs. In a LabVIEW implementation, the processing of event and time stamp data has to be implemented in hardware or the C/C++ driver program, which is needed so that LabVIEW can access the hardware on the PCI bus. In addition, an EPICS IOC could still be needed to correlate data across BPMs.

It must be noted that the performance of LabVIEW systems is highly dependent on the specific implementation; we exemplified this with regard to setpoints and reaction times. And while EPICS IOCs have a known, reasonable degradation with network load, where the system will stop responding to network requests but still perform local control, this is not possible with a LabVIEW/ActiveX approach because each COM call is a round-trip request. When the ActiveX CA server suffers from heavy network load, the LabVIEW program will degrade accordingly.

*Work supported by the Office of Basic Energy Science, Office of Science of the US Department of Energy, and by Oak Ridge National Laboratory.*